\documentclass[12pt,a4paper]{article}
\usepackage[top=3cm, bottom=3cm, inner=3cm, outer=3cm]{geometry}
\usepackage{helvet}

\usepackage[utf8]{inputenc}
\usepackage[english]{babel}
\usepackage{amsmath}
\usepackage{amsfonts}
\usepackage{amssymb}
\usepackage{graphicx}
\usepackage{mathtools}
\usepackage{rotating}
\usepackage{extarrows}
\usepackage{subcaption}
\usepackage{url}
\usepackage[superscript]{cite}
\makeatletter 
\renewcommand\@biblabel[1]{#1.} 
\makeatother

\usepackage{bm}
\def\bbeta{\bm{\beta}}
\def\bbetag{\bm{\beta}_{\mathrm{g}}}
\def\bgammag{\bm \gamma_{\mathrm{g}}}
\def\betag{\beta_{\mathrm{g}}}
\def\gammag{\gamma_{\mathrm{g}}}

\def\bx{\mathbf{x}}
\def\bg{\mathbf{g}}

\def\mC{\mathcal{C}}
\def\mM{\mathcal{M}}

\def\bx{\mathbf{x}}

\DeclareMathOperator{\E}{E}
\DeclareMathOperator{\T}{T}
\DeclareMathOperator{\I}{I}
\DeclareMathOperator{\Var}{Var}
\DeclareMathOperator{\Hmat}{H}
\DeclareMathOperator{\X}{X}
\DeclareMathOperator{\G}{G}
\DeclareMathOperator{\W}{W}
\DeclareMathOperator{\A}{A}

\newtheorem{definition}{Definition}[section]

\newtheorem{observation}{Observation}[section]
\newtheorem{assumption}{Assumption}[section]

\begin{document}
This is the peer reviewed version of the following article: Bjørnland, T, Bye, A, Ryeng, E, Wisløff, U, Langaas, M. Powerful extreme phenotype sampling designs and score tests for genetic association studies. Statistics in Medicine. 2018; 37: 4234– 4251., which has been published in final form at \url{https://doi.org/10.1002/sim.7914}. This article may be used for non-commercial purposes in accordance with Wiley Terms and Conditions for Use of Self-Archived Versions.

\title{Powerful extreme phenotype sampling designs and score tests for genetic association studies}
\date{}
\author{}
\maketitle

\begin{center}
THEA BJØRNLAND$^{1}$, ANJA BYE$^2$, EINAR RYENG$^3$, \\ ULRIK WISLØFF$^2$, METTE LANGAAS$^1$

\vspace{3mm}
\textit{$^1$ Department of Mathematical Sciences,
Norwegian University of Science and Technology, Trondheim,
Norway \\
$^2$ Department of Circulation and Medical Imaging, Norwegian University of Science and Technology, Trondheim, Norway \\
$^3$ Department of Cancer Research and Molecular Medicine, Norwegian University of Science and Technology, Trondheim, Norway}
\end{center}

\begin{abstract}
We consider cross-sectional genetic association studies (common and rare variants) where non-genetic information is available, or feasible to obtain for $N$ individuals, but where it is infeasible to genotype all $N$ individuals. We consider continuously measurable Gaussian traits (phenotypes). Genotyping $n<N$ extreme phenotype individuals can yield better power to detect phenotype-genotype associations, as compared to randomly selecting $n$ individuals. We define a person as having an extreme phenotype if the observed phenotype is above a specified threshold or below a specified thresholds. We consider a model where these thresholds can be tailored to each individual. The classical extreme sampling design is to set equal thresholds for all individuals. We introduce a design ($z$-extreme sampling) where personalized thresholds are defined based on the residuals of a regression model including only non-genetic (fully available) information. We derive score tests for the situation where only $n$ extremes are analyzed (complete case analysis), and for the situation where the non-genetic information on $N-n$ non-extremes is included in the analysis (all case analysis). For the classical design, all case analysis is generally more powerful than complete case analysis. For the $z$-extreme sample, we show that all case and complete case tests are equally powerful. Simulations and data analysis also show that $z$-extreme sampling is at least as powerful as the classical extreme sampling design and the classical design is shown to be at times less powerful than random sampling. The method of dichotomizing extreme phenotypes is also discussed.

\textit{Key words:} {GWAS, rare variants, outcome-dependent sampling, residual-based sampling, the HUNT study}
\end{abstract}

\section{Introduction}

Extreme phenotype sampling is a a two-stage design for genetic association studies, or other association studies where the covariate of interest is expensive or infeasible to obtain for all individuals. Selective genotyping of extreme-phenotype individuals has been proposed as a strategy for achieving good statistical power under sample size limitations \cite{lebowitz1987trait, lander1989mapping, darvasi1992selective ,slatkin1999disequilibrium, chen2005linkage, wallace2006improved, van2000power, xing2009power}. Inexpensive `non-genetic' information (phenotype/trait measurements, age, sex, lifestyle variables, etc.) is collected on $N$ individuals in the first stage. In the second stage, $n < N$ individuals are selected for genotyping. Criteria for selection are based on stage 1 observations with the intention to increase the power to detect genotype-phenotype associations, as compared to randomly selecting $n$ individuals. For simple linear regression models, Huang and Lin proposed likelihood methods for extreme samples that made full use of the genetic data, that modelled the continuous rather than a discretized phenotype and properly accounted for the selective genotyping design \cite{huang2007eps}. Tao \textit{et al} developed methods for parameter estimation and a Wald test for two-stage sampling designs \cite{tao2017}. Derkach \textit{et al} derived score tests for similar designs \cite{derkach2015score}. Extreme phenotype sampling has also been proposed for studies of rare genetic variants \cite{li2011eps, guey2011power, barnett2013rareeps, tao2015epsmultivar}. 

Consider a common biallelic genetic variant $\text{g}$ (SNP), an additive genetic model, a Gaussian phenotype $y$ and the model $y = \beta_0 + \betag \text{g} + \varepsilon$, where $\varepsilon \sim \text{N}(0,\sigma^2)$. The genetic covariate $\text{g}$ takes values 0 (zero minor alleles), 1 (one minor allele) or 2 (two minor alleles). To establish whether $\betag \neq 0$, a powerful sample will have a high proportion of individuals with genotypes $0$ and $2$. Assuming Hardy-Weinberg equilibrium, if the minor allele frequency of a SNP is 0.25, then in any random sample from the population approximately $56\%$ of individuals have genotype 0, $37\%$ will have genotype 1, and only $6\%$ will have genotype 2. If indeed the genetic variant has an additive effect on the phenotype, and there are no other covariates influencing the phenotype, then in one end of the empirical phenotype distribution we will find an increased proportion of subjects with genotype 2, while in the other end we will mainly find subjects with genotype 0. The extreme sample can therefore give better power than random sampling. For studies of rare genetic variants (minor allele frequency below $5\%$) an extreme sample should have a higher proportion of individuals with at least one minor allele if the genetic variant is causal.

If the phenotype is influenced by other `non-genetic' covariates $x$ (e.g. age, sex, lifestyle), $y = \beta_0 + \beta_1 x + \betag \text{g}  + \varepsilon$, then it is not given that the phenotypic extremes will have a more favorable distribution of genotypes than a random sample. Therefore, we consider a design where extremes are defined based on a `residual phenotype' ($z = y - \widehat{\beta}_0 - \widehat{\beta}_1 x$), i.e. the observed phenotype after adjustment for relevant non-genetic covariates. We refer to this design as the $z$-extreme sampling design. 

Define \textit{complete cases} as the $n<N$ individuals that are genotyped. Under extreme phenotype sampling, the most straightforward association analysis for the complete cases is to compare the two extreme groups using models for binary outcomes \cite{van2000power, peloso2016phenotypic}. This approach has been shown to be less powerful than other methods \cite{huang2007eps}. A complete case test should rather be derived based on the probability distribution of the extremes \cite{huang2007eps, tang2010equivalence}. Analysis of \textit{all cases} ($n$ complete cases and $N-n$ cases with missing genotype information) is based on methods for missing at random covariates \cite{derkach2015score, tao2017}. When there are no non-genetic covariates in the model, complete case and all case analysis are equally powerful \cite{huang2007eps,derkach2015score}. Otherwise, all case analysis is generally more powerful than complete case analysis. We prove that when extremes are defined based on the residual phenotype ($z$-extreme sampling), the complete case and all case score tests are equally powerful, regardless of the impact of non-genetic covariates.

When the sampling design is based on the construction of the residual phenotype $z$, it is not unnatural to consider testing for an association directly with this adjusted phenotype. Although these residuals by design are not independent, we show that score tests `naively' applied to residual phenotype models ($z = \gamma_0 + \gammag \text{g} + \varepsilon$) will control the type-1 error rate at the desired significance level. If the genetic covariate is correlated with some of the non-genetic covariates in the model ($x$) then such residual phenotype tests are shown to be less powerful than tests based on the original regression model ($y = \beta_0 + \beta_1 x + \betag \text{g}  + \varepsilon$).

In Section 2 of this paper we give some statistical background and specific results for random sampling. In Section 3 we present score tests for testing association with one common genetic variant in extreme samples. The multivariate version of this theory is given in the Supplementary Material. In Section 4 we consider tests of rare genetic variants. We derive association tests similar to the SKAT test\cite{wu2011SKAT, lee2012SKATO} for rare genetic variants under extreme sampling. In Section 5 we illustrate the results of Sections 3 and 4 by simulations, in Section 6 we apply methods to data from the HUNT study (Helseundersøkelsen i Nord-Trøndelag) \cite{HUNT}, and in Section 7 we discuss our findings.

\section{Statistical background}

We consider the linear model 
\begin{align}
\mathbf{Y} = \X \bbeta + \bg \betag  + \bm \varepsilon, \quad \bm \varepsilon \sim \text{MVN}(\bm 0, \sigma^2 \I),
\label{eq:lmysimple}
\end{align}
where $\mathbf{Y}$ is an $N$-vector of phenotype observations, $\X$ an $N \times (d+1)$ matrix of the intercept and $d$ non-genetic covariates ($d+1 \ll N$), $\bg$ an $N$-vector of genotype observations and $\I$ is the $N \times N$ identity matrix. The residual vector $\bm \varepsilon$ is multivariate normal (MVN) with mean $\bm 0$ and covariance matrix $\sigma^2 \I$.  We test $H_0: \betag = 0$ against $H_1: \betag \neq 0$ (genotype-phenotype association) using score tests. We further generalize model \eqref{eq:lmysimple} to
\begin{align}
\mathbf{Y} = \X \bbeta + \G \bbetag  + \bm \varepsilon, \quad \bm \varepsilon \sim \text{MVN}(\bm 0, \sigma^2 \I),
\label{eq:lmygeneral}
\end{align}
where the $N \times \nu$ genotype matrix $\G$ represents a selection of $\nu$ genetic variants ($\nu \ll N$), and derive score tests for testing $H_0: \bbetag = \bm 0$ against $H_1: \bbetag \neq \bm 0$. 

\begin{definition}[The score test for $H_0: \bbetag = \bm 0$]
\label{def:scoretest}
Consider a log likelihood function $l(\bbetag, \bm \theta)$, where $\bm \theta$ is a vector of `nuisance parameters' and $\bbetag$ are the parameters of interest. When $H_0$ is assumed true ($\bbetag = \bm 0$) the maximum likelihood estimators of nuisance parameters, denoted $\widehat{\bm \theta}$, are found by solving $\partial l / \partial \bm \theta(\bbetag = \bm 0) = \bm 0$. Define the score statistic
$$\mathbf{U}_{\bbetag} = \frac{\partial l }{\partial \bbetag} \left(\bbetag = \bm 0, \bm \theta = \widehat{\bm \theta}\right),$$ and the information matrix $$ \mathbf{F}(\bbetag, \bm \theta) = \begin{bmatrix}
\mathbf{F}_{\bbetag \bbetag}(\bbetag, \bm \theta) & \mathbf{F}_{\bbetag \bm \theta}(\bbetag, \bm \theta) \\ \mathbf{F}_{\bm \theta \bbetag}(\bbetag, \bm \theta) & \mathbf{F}_{\bm \theta \bm \theta}(\bbetag, \bm \theta)
\end{bmatrix} =  - \E  \begin{bmatrix}
\frac{\partial^2 l}{\partial \bbetag \partial \bbetag^{\T}} & \frac{\partial^2 l}{\partial \bbetag \partial \bm \theta^{\T}} \\ \frac{\partial^2 l}{\partial \bm \theta \partial \bbetag^{\T}} & \frac{\partial^2 l}{\partial \bm \theta \partial \bm \theta^{\T}}
\end{bmatrix}. $$ The variance of the score statistic is $$\Var(\mathbf{U}_{\bbetag}) = \mathbf{F}_{\bbetag \bbetag}(\bm 0, \widehat{\bm \theta}) - \mathbf{F}_{\bbetag \bm \theta}(\bm 0, \widehat{\bm \theta})\mathbf{F}_{\bm \theta \bm \theta}^{-1}(\bm 0, \widehat{\bm \theta}) \mathbf{F}_{\bm \theta \bm \bbetag}(\bm 0, \widehat{\bm \theta}).$$ The score test statistic $T = \mathbf{U}_{\bbetag}^{\T}\Var(\mathbf{U}_{\bbetag})^{-1}\mathbf{U}_{\bbetag}$ is asymptotically $\chi^2_{\nu}$-distributed under the null. When $\betag$ is a scalar then $\nu = 1$ and $T = U_{\betag}^2/\Var(U_{\betag})$.
\end{definition}

\begin{definition}[Residual phenotype]
\label{def:res_pheno}
Consider the null model $\mathbf{Y} = \X \bm \beta + \bm \varepsilon$, and let $\widehat{\bm \beta} =  (\X^{\T}\X)^{-1}\X^{\T} \mathbf{Y}$ be the maximum likelihood estimator for $\bbeta$. Define the residual phenotype as $\mathbf{Z} = \mathbf{Y} - \widehat{\mathbf{Y}} = \mathbf{Y} - \X \widehat{\bm \beta} = (\I - \Hmat)\mathbf{Y}$, where $\Hmat = \X (\X^{\T}\X)^{-1}\X^{\T}$.
\end{definition}
If the null hypothesis is true then $\mathbf{Z} \sim \text{MVN}(\bm 0, \sigma^2(\I - \Hmat))$. The residual phenotypes $Z_1, \ldots, Z_N$ are therefore generally not i.i.d. The analogs to models \eqref{eq:lmysimple} and \eqref{eq:lmygeneral} are the \textit{residual phenotype models}
\begin{align}
\mathbf{Z} = \mathbf{1}\gamma_0 + \bg \gammag + \bm \varepsilon_z, \quad \text{ and } \quad \mathbf{Z} = \mathbf{1}\gamma_0 + \G \bgammag + \bm \varepsilon_z.
\label{eq:lmz}
\end{align}
The parameter $\gammag$ ($\bgammag$) can be interpreted as a surrogate for $\betag$ ($\bbetag$) but we assume throughout that the main interest lies in testing $H_0: \betag = 0$ ($H_0: \bbetag = \bm 0$).

\subsection{Score tests for the linear regression model}
\label{section:ScoreRS}

In this section only, assume that $(Y_i, \bx_i, \text{g}_i)$ is observed for all $N$ individuals, randomly sampled from some (infinite) population of interest. The log likelihood function is $l = -N \log(\sigma) - \frac{1}{\sigma^2}\sum_{i = 1}^N (Y_i  -\bx_i^{\T}\bbeta - \text{g}_i \betag)^2$. The score test statistic (Definition \ref{def:scoretest}) for testing $H_0: \betag = 0$ in model \eqref{eq:lmysimple} is given by $T = U_{\betag}^2 /\Var(U_{\betag})$, where 
\begin{align}
U_{\betag} = \frac{1}{\widehat{\sigma}^2} \bg^{\T}(\I - \Hmat)\mathbf{Y}, \quad \Var(U_{\betag}) = \frac{1}{\widehat{\sigma}^2} \bg^{\T}(\I - \Hmat)\bg.
\label{eq:scoreRS}
\end{align}
The hat matrix is defined under the null by $\Hmat = \X(\X^{\T}\X)^{-1}\X^{\T}$, and the maximum likelihood estimator for $\sigma^2$ is $\widehat{\sigma}^2 = \frac{1}{N} \mathbf{Y}^{\T}(\I - \Hmat)\mathbf{Y}$. The test statistic $T$ has an asymptotic $\chi^2_{1}$ distribution when the null hypothesis is true, and the null hypothesis is rejected when $T>t_{\alpha}$, where $t_{\alpha}$ is a critical value from the  $\chi^2_{1}$ distribution. `Naively' applying the same test to the residual phenotype model \eqref{eq:lmz}, we obtain $T_z = U_{\gammag}^2/\Var(U_{\gammag})$, where
\begin{align}
U_{\gammag} = \frac{1}{\widehat{\sigma}_z^2} \bg^{\T}(\I - \frac{1}{N}\mathbf{1}\mathbf{1}^{\T})\mathbf{Z}, \quad \Var(U_{\gammag}) = \frac{1}{\widehat{\sigma}_z^2} \bg^{\T}(\I - \frac{1}{N}\mathbf{1}\mathbf{1}^{\T})\bg.
\label{eq:scoreRSz}
\end{align}
Note that since the assumption $\bm \varepsilon_z \sim \text{MVN}(\bm 0, \sigma^2 \I)$ is generally incorrect, the asymptotic null distribution of $T_z$ is not necessarily $\chi^2_1$. However, an asymptotically valid test for $H_0: \betag = 0$ against $H_1: \betag \neq 0$ in model \eqref{eq:lmysimple} when is to reject $H_0$ whenever $T_z > t_{\alpha}$, where $t_{\alpha}$ is a critical value from the $\chi^2_1$-distribution. By \textit{valid} we mean that the type-1 error rate is controlled at the desired level $\alpha$; $\Pr(\text{reject } H_0|H_0 \text{ true}) \leq \alpha$. The residual phenotype test is valid because $T_z \leq T$ (proof given in Supplementary Material Section 2.1), so that $T_z < t_{\alpha}$ whenever $T< t_{\alpha}$. Rejecting $H_0$ whenever $T_z > t_{\alpha}$ is then generally a less powerful test for $H_0: \betag = 0$ because the event $T_z > t_{\alpha}$ will occur less often than $T > t_{\alpha}$. When $\X$ and $\bg$ are \textit{uncorrelated} (sample correlation between each column of $\X$ and $\bg$ is zero, $(\X-\frac{1}{N}\mathbf{1}\mathbf{1}^{\T}\X)^{\T}(\bg-\frac{1}{N}\mathbf{1}\mathbf{1}^{\T}\bg) = \bm 0$, then $T_z = T$. This result also generalizes to the score test for $H_0: \bbetag = \bm 0$ (model \eqref{eq:lmygeneral}) where the test statistics are compared to the $\chi^2_{\nu}$-distribution, see Supplementary Material Section 2.

\subsection{Missing covariates}
\label{section:MissingRS}
Consider a sample of size $N$ (randomly generated from the infinite population of interest) where the genetic covariate is missing for $N-n$ individuals, while other non-genetic covariates and the response variable are observed for all $N$ individuals.

\begin{definition}[Complete case analysis]
\label{def:CC}
Only the $n$ individuals with observed genotype ($\textsl{g}_i$), non-genetic covariates ($\bx_i$) and phenotype ($y_i$) are analyzed. The set of indexes $\{i: i \in \mC \}$ denotes all completely observed individuals. Let $\bg_{\mC}$ be the $n$-vector of observed genotypes, $\mathbf{Y}_{\mC}$ the complete case phenotype vector and $\X_{\mC}$ the complete case covariate matrix.
\end{definition}

\begin{definition}[All case analysis]
\label{def:AC}
For all case analysis we include the $n$ complete case observations $(y_i, \bx_i, \textsl{g}_i)$ for all $i \in \mC$ and the $N-n$ observations of $(y_i, \bx_i)$ for all $i \in \mM$, where $\mM = \{i \in \{1, \ldots, N\}: i \not \in \mC \}$.
\end{definition}

Let $\mathbf{R}$ be a random $N$-vector that indicates missing entries in the genetic covariate vector $\bg$. The missing-mechanism is \textit{missing completely at random} (MCAR) if $\Pr(\mathbf{R}= \mathbf{r}|\mathbf{Y}, \X, \bg) = \Pr(\mathbf{R}= \mathbf{r})$ and \textit{missing at random} (MAR) if $\Pr(\mathbf{R}= \mathbf{r}|\mathbf{Y}, \X, \bg) = \Pr(\mathbf{R}= \mathbf{r}|\mathbf{Y}, \X)$ \cite{littlerubin2002}. In the case of MCAR covariates, complete case analysis is performed by assuming that the complete case sample ($\mathbf{Y}_{\mC}$, $\X_{\mC}$ and $\bg_{\mC}$) is a random sample from the (infinite) population of interest. All case analysis is performed by imputing the mean genotype where genotypes are missing. The mean is estimated and imputed separately for genetically different subgroups. For sufficiently large samples, complete case and all case score tests are approximately equally powerful (Supplementary Material Section 5.5).

\subsection{Extreme phenotype sampling}

We define a general extreme phenotype sampling design where the classification rule (extreme or not extreme) can be tailored to each individual in the sample. 

\begin{definition}[Extreme phenotype sampling]
\label{def:extreme_pheno}
Individual $i$ has an extreme phenotype if $Y_i < l_i$ or $Y_i > u_i$, where $l_i$ and $u_i$ are known thresholds. All individuals who are classified as extreme are selected for genotyping.
\end{definition}
Two special cases of this design are of particular interest:

\begin{definition}[$y$-extreme sampling]
\label{def:y_extreme}
Individuals are $y$-extreme if $Y_i < l$ or \\ $Y_i > u$, where thresholds $l$ and $u$ are equal for all $i \in \{1, \ldots, N\}$.
\end{definition}

\begin{definition}[$z$-extreme sampling]
\label{def:z_extreme}
Individuals are $z$-extreme if $Z_i < l_z$ or \\ $Z_i > u_z$, where thresholds $l_z$ and $u_z$ are equal for all $i \in \{1, \ldots, N\}$, and $Z_i$ is the residual phenotype (Definition \ref{def:res_pheno}). By Definition \ref{def:extreme_pheno}, individual $i$ has an extreme phenotype if $Y_i < l_i$ or $Y_i > u_i$, where $l_i = l_z + \bx_i^{\T}\widehat{\bbeta}$ and $u_i = u_z + \bx_i^{\T}\widehat{\bbeta}$.
\end{definition}

\section{Single variant tests for extreme samples}
\label{section:EPSsingle}

We consider model \eqref{eq:lmysimple} under extreme phenotype sampling (Definition \ref{def:extreme_pheno}) and score tests for $H_0: \betag = 0$ vs $H_1: \betag \neq 0$.

\subsection{Complete case analysis}

Using the conditional phenotype distribution $Y_i | (Y_i < l_i \cup Y_i > u_i)$, where classification rules $l_i$ and $u_i$ are determined before seeing the data, the likelihood for the complete cases (Definition \ref{def:CC}) is
\begin{align*}
L_{\mC} = \prod_{i \in \mC} \frac{\frac{1}{\sigma} \phi\left( \frac{Y_i - \mu_i}{\sigma} \right)}{1 - \Phi\left( \frac{u_i - \mu_i}{\sigma} \right) + \Phi\left( \frac{l_i - \mu_i}{\sigma} \right)},
\end{align*}
where $\phi$ is the density function and $\Phi$ is the cumulative probability distribution for the standard normal distribution, and $\mu_i = \bx_i^{\T}\bbeta + \text{g}_i \betag$. Define (similarly to Tang \cite{tang2010equivalence}),
\begin{align}
h_{ij} = \frac{- \phi\left( \frac{u_i - \mu_i}{\sigma} \right) \cdot \left( \frac{u_i - \mu_i}{\sigma} \right)^j +  \phi\left( \frac{l_i - \mu_i}{\sigma} \right) \cdot \left( \frac{l_i - \mu_i}{\sigma} \right)^j}{1 - \Phi\left( \frac{u_i - \mu_i}{\sigma} \right) + \Phi\left( \frac{l_i - \mu_i}{\sigma} \right)},
\label{eq:hfun}
\end{align}
\begin{align*}
a_{i} = 1 - h_{i1} - h_{i0}^2, \quad b_i = -h_{i0} - h_{i2} - h_{i0}h_{i1}, \quad c_i = 2 - h_{i1} - h_{i3} - h_{i1}^2,
\end{align*}
and let $\mathbf{h}_j$, $\mathbf{a}$, $\mathbf{b}$ and $\mathbf{c}$ be the corresponding $n$-vectors for these expressions. Define $\A = \text{Diag}(\mathbf{a})$ and $\mathbf{W}_{\mC} = \mathbf{Y}_{\mC} + \sigma \mathbf{h}_{0}$. Let complete case maximum likelihood estimators of $\bbeta$ and $\sigma$ under the null ($\betag = 0$) be denoted by a tilde $\sim$, and let $\widetilde{\mathbf{h}}_j$, $\widetilde{\mathbf{W}}_{\mC}$, $\widetilde{\A}$, $\widetilde{\mathbf{b}}$ and $\widetilde{\mathbf{c}}$ be the expressions $\mathbf{h}_j$, $\mathbf{W}_{\mC}$, $\A$, $\mathbf{b}$ and $\mathbf{c}$ evaluated at $\widetilde{\sigma}$, $\widetilde{\bbeta}$ and $\betag = 0$. The complete case score test statistic is defined by $T_{\mC} = U_{\mC}^2/\Var(U_{\mC})$, where
\begin{align}
U_{\mC} = \frac{1}{\widetilde{\sigma}^2} \bg_{\mC}^{\T}(\I-\Hmat_{\mC})\widetilde{\mathbf{W}}_{\mC}, \quad \Hmat_{\mC} = \X_{\mC}(\X_{\mC}^{\T}\X_{\mC})^{-1}\X_{\mC}^{\T},
\label{eq:syc}
\end{align}
and
\begin{align}
\Var(U_{\mC} ) = \frac{1}{\widetilde{\sigma}^2}\left( \bg_{\mC}^{\T} \widetilde{\A} \bg_{\mC} - 
\bg_{\mC}^{\T} 
\begin{bmatrix}
\widetilde{\A} \X_{\mC} & 
\widetilde{\mathbf{b}}
\end{bmatrix}
\begin{bmatrix}
\X_{\mC}^{\T}  \widetilde{\A} \X_{\mC} & \X_{\mC}^{\T} \widetilde{\mathbf{b}} \\
\widetilde{\mathbf{b}}^{\T}\X_{\mC} & \mathbf{1}^{\T}\widetilde{\mathbf{c}}
\end{bmatrix}^{-1}
\begin{bmatrix}
\X_{\mC}^{\T} \widetilde{\A} \\ 
\widetilde{\mathbf{b}}^{\T}
\end{bmatrix} \bg_{\mC}
  \right).
 \label{eq:varsyc}
\end{align}
Details are given in Supplementary Material Section 4. When the null hypothesis is true, the test statistic $T_{\mC}$ has an asymptotic $\chi^2_{1}$ distribution.

\subsubsection{$y$-extreme sampling}
The score test statistic for the $y$-extreme sample follows directly from the above by inserting $l_i = l $ and $u_i = u$ into the functions $h_{ij}$ \eqref{eq:hfun}.  This leads to no simplifications unless $\X_{\mC} = \mathbf{1}_n$. Then, the score test statistic $T_{\mC}$ is equal to the test statistic one would find by assuming that the complete case sample ($\{i: i \in \mC\}$) is a random sample \cite{tang2010equivalence}. Indeed, if the null hypothesis is true and there are no non-genetic covariates influencing the phenotype, then $Y_i \sim \text{N}(\beta_0, \sigma^2)$  $\forall i$, and all individuals have equal probability to be sampled for genotyping since $\Pr(Y_i < l \cup Y_i > u) = \Pr(Y_j < l \cup Y_j > u)$ for all $i , j$.

\subsubsection{$z$-extreme sampling}

Under $z$-extreme sampling (Definition \ref{def:z_extreme}) the expression for the complete case score test statistic can be simplified. Strictly, $\widehat{\bbeta}$ should be estimated from a pilot study so that the $z$-extreme sampling rule is determined before seeing data. However, we may assume that $\widehat{\bbeta}$ estimated from the full sample (large $N$) is unbiased and estimated with low variance and may therefore be regarded as deterministic. Recall that $\widetilde{\bbeta}$ is the maximum likelihood estimator based on the complete cases. For large $N$ and $n$, $u_i - \widetilde{\mu}_i = u_z + \bx_i^{\T}\widehat{\bbeta} - \bx_i^{\T}\widetilde{\bbeta} \approx u_z$, $l_i - \widetilde{\mu}_i = l_z + \bx_i^{\T}\widehat{\bbeta} - \bx_i^{\T}\widetilde{\bbeta} \approx l_z$, and $\widetilde{h}_{ij}$, $\widetilde{a}_i$, $\widetilde{b}_i$, $\widetilde{c}_i$ are (approximately) equal for all $i$. Equations \eqref{eq:syc} and \eqref{eq:varsyc} simplify to
\begin{align}
U_{\mC} \approx \frac{1}{\widetilde{\sigma}^2} \bg_{\mC}^{\T}(\I - \Hmat_{\mC})\mathbf{Y}_{\mC}, \quad \Var(U_{\mC} ) \approx \frac{\widetilde{a}}{\widetilde{\sigma}^2} \bg_{\mC}^{\T}\left(\I - \Hmat_{\mC} \right) \bg_{\mC}.
\label{eq:szc}
\end{align}

\begin{observation}
\label{obs:CCRS}
Under $z$-extreme sampling $T_{\mC} \approx T_n$, where $T_n$ is the test statistic obtained by assuming that the complete case sample is a random sample; $T_n = U_n^2/\Var(U_n)$, where $U_n = \frac{1}{\widehat{\sigma}_n^2}\bg_{\mC}^{\T}(\I - \Hmat_{\mC})\mathbf{Y}_{\mC}$, $\Var(U_n) = \frac{1}{\widehat{\sigma}_n^2}\bg_{\mC}^{\T}(\I - \Hmat_{\mC})\bg_{\mC}$, and $\widehat{\sigma}_n^2 = \frac{1}{n}\mathbf{Y}^{\T}(\I - \Hmat_{\mC})\mathbf{Y}_{\mC}$.
\end{observation}
Observation \ref{obs:CCRS} follows from the fact that $\widetilde{\sigma}^2 = \frac{\widehat{\sigma}_n^2}{\widetilde{a}}$, see Supplementary Material Section 4.4.2, and can be seen as a generalization of the result by Tang \cite{tang2010equivalence}. If the null hypothesis is true, then $Y_i - \X_i \bbeta \sim \text{N}(0, \sigma^2)$ $\forall i$, and $$\Pr((Y_i - \bx_i^{\T} \bbeta  < l_z) \cup (Y_i - \bx_i^{\T} \bbeta  > u_z)) = \Pr((Y_j - \bx_j^{\T} \bbeta  < l_z) \cup (Y_j - \bx_j^{\T} \bbeta  > u_z))$$ for all  $i , j$. Replacing $\bbeta$ by $\widehat{\bbeta}$, each individual in the sample will under $H_0$ have approximately equal probability to be sampled for genotyping.

\paragraph{Residual phenotype test}

If, under $z$-extreme sampling, we derive a test directly from the residual phenotype model \eqref{eq:lmz}, then we obtain
\begin{align}
U_{z_{\mC}} = \frac{1}{\widetilde{\sigma}_z^2} \bg_{\mC}^{\T}(\I-\frac{1}{n}\mathbf{1}\mathbf{1}^{\T})\mathbf{Z}_{\mC}, \quad \Var(U_{z_\mC}) = \frac{\widetilde{a}_z}{\widetilde{\sigma}_z^2}  \bg_{\mC}^{\T} (\I - \frac{1}{n}\mathbf{1}\mathbf{1}^{\T}) \bg_{\mC},
\label{eq:uzc}
\end{align}
and $T_{z_{\mC}} = U_{z_{\mC}}^2/\Var(U_{z_{\mC}})$. Since the model assumption $\bm \varepsilon_z \sim \text{MVN}(\bm 0, \sigma^2 \I)$ is generally incorrect (see Definition \ref{def:res_pheno}), the asymptotic null distribution of $T_{z_{\mC}}$ is not necessarily $\chi^2_1$. Observation \ref{obs:asympt_zCC} is proven in Supplementary Material Section 4.4.3.
\begin{observation}
\label{obs:asympt_zCC}
Under $z$-extreme sampling, $T_{z_{\mC}} \approx T_{{\mC}}$ if  $\X_{\mC}$ is column-wise uncorrelated with $\bg_{\mC}$. Otherwise, $T_{z_{\mC}} \leq T_{{\mC}}$. If $t_{\alpha}$ is a critical value from the $\chi^2_{1}$-distribution, then rejecting $H_0$ when $T_{z_{\mC}} > t_{\alpha}$ is a valid but conservative test.
\end{observation}

\subsection{All case analysis}
\label{section:AC}

Assuming $y_i$ and $\bx_i$ are observed for all $N$ individuals, the extreme sampling design (including $y$-extreme and $z$-extreme sampling) satisfies the MAR condition. For all case analysis (Definition \ref{def:AC}) we use a likelihood model for a sample with MAR covariates \cite{littlerubin2002} based on the following assumptions:

\begin{assumption}
\label{ass:1}
$G$ is a discrete random variable (genotype) with sample space $\{g_k; k = 1, \ldots, K\}$.
\end{assumption}

\begin{assumption}
\label{ass:2}
The distribution of $G$ (genotype frequencies) can differ between different groups of individuals. Assume that there are $J$ groups and let $\Pr(G = g_k|j) = p_{jk}$, $k = 1, \ldots, K$, be the point probabilities of $G$ in group $j$, $j = 1, \ldots, J$. Groups are defined by unique covariate patterns of $\X^*$ (dummy coded categorical covariates), where $\X^*$ denotes $d^* +1 \leq d +1$ columns of $\X$ (including intercept). 
\end{assumption}
Let $G_i$ denote the random genotype for individual $i$. For individuals $i \in \mC$, $\text{g}_i$ denotes the actual observation of $G_i$. Assumption \ref{ass:2} implies that genetic and non-genetic covariates $\bg$ and $\X$ can be correlated. Generally, $\X$ can contain continuous covariates, but $\X^*$ is discrete (possibly including discretized and dummy coded continuous covariates). Let $\mC_j$ denote the set of all $n_j$ genotyped individuals $i \in \mC$ who are also in group $j$, let $\mM_j$ denote the set of all $m_j$ missing-genotype individuals $i \in \mM$ who are also in group $j$, and let $N_j = n_j + m_j$, $\sum_j n_j = n$, $\sum_j m_j = m$, $\sum_j N_j = N$  and $n+m = N$. Let $n_{jk}$ be the random variable that counts observations of genotype $g_k$ among genotyped individuals in group $j$ ($i \in \mC_j$). The all case likelihood is
\begin{align*}
L = & \prod_{i \in \mC} f_{Y_i|G_i = \text{g}_i}(y_i;\bx_i) \Pr(G_i = \text{g}_i; \bx_i) \prod_{i \in \mM}\sum_{k} f_{Y_i|G_i = g_k}(y_i; \bx_i) \Pr(G_i = g_k; \bx_i)\\
\propto &  \prod_{i \in \mC} \frac{1}{\sigma} \phi \left( \frac{Y_i - \mu_i}{\sigma} \right) \prod_j \prod_k p_{jk}^{n_{jk}} \prod_j \prod_{i \in \mM_j} \sum_k \frac{1}{\sigma} \phi \left( \frac{Y_i - \mu_i(g_k)}{\sigma} \right) p_{jk},
\end{align*}
where $\mu_i = \bx_i^{\T} \bbeta + \text{g}_i \betag$ for $i \in \mC$ and $\mu_i(g_k) = \bx_i^{\T} \bbeta + g_k \betag$ for $i \in \mM$. The maximum likelihood estimators of $\bbeta$, $\sigma$ and $p_{jk}$ under the null are
\begin{align}
\widehat{\bbeta} = \left(\X^{\T} \X\right)^{-1} \X^{\T} \mathbf{Y}, \quad \widehat{\sigma}^2 = \frac{1}{N} \mathbf{Y}(\I-\Hmat)\mathbf{Y}, \quad \widehat{p}_{jk} = n_{jk}^{\text{obs}}/n_{j},
\label{eq:mleACy}
\end{align}
and the estimated mean under the null is $\widehat{\bm \mu} = \X \widehat{\bbeta} = \Hmat \mathbf{Y}$. These closed-form expressions make model fitting under the null computationally efficient for all case analysis. The score test for the all case likelihood was derived by Derkach \textit{et al} \cite{derkach2015score} for models where $\X^* = \X$ (Assumption \ref{ass:2}). Our test is derived using a slightly different approach (see Supplementary Material Section 5), but the result is equivalent when $\X^* = \X$ and probabilities $p_{jk}$ are unknown. Let $\bg_M$ be an $N$-vector where for individuals $i \in \mC$, $\bg_{M,i} = \text{g}_i$, while for $i \in \mM_j$ the mean genotype across individuals $i \in \mC_j$ is imputed. The all case score test statistic is $T_A = U_A^2/\Var(U_A)$, where
\begin{align}
U_A = \frac{1}{\widehat{\sigma}^2} \bg_M^{\T}(\I-\Hmat) \mathbf{Y},
\label{eq:sy}
\end{align}
and
\begin{align}
& \Var(\mathbf{U}_A)
 = \frac{1}{\sigma^2}\bg_M^{\T}(\I - \Hmat)\bg_M - \frac{1}{\sigma^2} \sum_j n_j \Var(G|j) \nonumber \\
& + \frac{1}{\sigma^4} \sum_j \Var(G|j) \left( \E \sum_{i \in \mC_j}(Y_i - \mu_i(\betag = 0))^2 - \frac{1}{n_j} \left( \E \sum_{i \in \mC_j}(Y_i - \mu_i(\betag = 0)) \right)^2 \right) \label{eq:varianceAllCase} \\
& = \frac{1}{\sigma^2}\bg_M^{\T}(\I - \Hmat)\bg_M - \frac{1}{\sigma^2} \sum_j \Var(G|j) \left( \sum_{i \in \mC_j}h_{i1}(\betag = 0) + \frac{1}{n_j} \left(  \sum_{i \in \mC_j} h_{i0}(\betag = 0) \right)^2 \right),
\label{eq:varianceAllCase3}
\end{align} 
where $h_{i0}$ and $h_{i1}$ are defined in Equation \eqref{eq:hfun}.
The score test statistic has an asymptotic $\chi^2_1$ distribution under the null.

\subsubsection{$y$-extreme sampling}
When there are no non-genetic covariates $\X$, and therefore no groups (Assumption \ref{ass:2}), then $U_A \approx \frac{1}{\widehat{\sigma}^2} \bg_{\mC}^{\T}(\I-\Hmat_{\mC}) \mathbf{Y}_{\mC}$ and $\Var(U_A) \approx \frac{\widehat{a}}{\widehat{\sigma}^2}\bg_{\mC}^{\T}(\I-\Hmat_{\mC}) \bg_{\mC}$, where $\widehat{a} = a(\widehat{\sigma})$, see Equation \eqref{eq:hfun}. Then, the complete case and all case score test statistics are approximately equal. Details are given in Supplementary material Section 5.6.

\subsubsection{$z$-extreme sampling}
Under $z$-extreme sampling (Definition \ref{def:z_extreme}) the all case score statistic \eqref{eq:sy} can be approximated by
\begin{align*}
U_A \approx \frac{1}{\widehat{\sigma}^2} \bg_{\mC}^{\T} (\I - \Hmat_{\mC}) \mathbf{Y}_{\mC}.
\end{align*}
The variance \eqref{eq:varianceAllCase} can be approximated by
\begin{align*}
\Var(U_A) \approx \frac{\widehat{a}}{\widehat{\sigma}^2}\bg_{\mC}^{\T}(\I - \Hmat_{\mC})\bg_{\mC},
\end{align*}
where $a$ is given in Equation \eqref{eq:hfun}, and $\widehat{a}$ denotes that the function is evaluated at the all case maximum likelihood estimator for $\sigma$. Details are given in Supplementary Material Section 5.7.

\begin{observation}
\label{obs:z_extreme_all_complete}
Under $z$-extreme sampling, $U_{A} \approx \frac{\widetilde{\sigma}^2}{\widehat{\sigma}^2} U_{\mC} \approx U_{\mC}$, where $U_{\mC}$ is the compete case score statistic  \eqref{eq:syc}, $U_A$ is the all case score statistic \eqref{eq:sy}, $\widetilde{\sigma}$ is the complete case maximum likelihood estimator of $\sigma$ and $\widehat{\sigma}$ is the all case estimator. Also, $\Var(U_{A}) \approx \frac{\widetilde{\sigma}^2}{\widehat{\sigma}^2} \frac{\widehat{a}}{\widetilde{a}} \Var(U_{\mC}) \approx \Var(U_{\mC})$. Under $z$-extreme sampling, the all case and complete case score tests are therefore approximately equal. 
\end{observation}

\paragraph{Residual phenotype test} Under $z$-extreme sampling it follows from Observation \ref{obs:z_extreme_all_complete} that also the residual phenotype all case score test statistic $T_{z_{A}}$ is approximately equal to the residual phenotype complete case score test $T_{z_{\mC}}$ \eqref{eq:uzc}, see Supplementary Material Section 5.7.1.

\section{Rare variant tests for extreme samples}
\label{sec:rarevar}
In Supplementary Material Sections 4 and 5 we show that all results from the previous section generalize to testing $H_0: \bbetag = \bm 0$ in model \eqref{eq:lmygeneral} (i.e. simultaneously testing $\nu$ genetic variants, $1 < \nu \ll N$). Here we consider testing $\nu$ rare genetic variants and we derive two rare variant association tests for the extreme phenotype sampling design; the `collapsing test' \cite{morgenthaler2007collapse, li2008collapse, madsen2009collapse, morris2010collapse} and  the `variance component test' \cite{wu2011SKAT, lee2012SKATO}. Under extreme sampling, a binary response collapsing test was used by Peloso \textit{et al} \cite{peloso2016phenotypic}. A complete case collapsing test was studied by Li \textit{et al} \cite{li2011eps}. A variance component test (the SKAT test) for complete case analysis of $y$-extreme samples was proposed by Barnett \textit{et al} \cite{barnett2013rareeps}. In the extreme sampling complete case setting, we find that estimation of the variance parameter $\sigma^2$ must be accounted for when deriving the score test (the parameters $\sigma$ and $\bbetag$ are not orthogonal ($\mathbf{F}_{\sigma \bbetag} \neq 0$), see Supplementary Material Section 4.2). We therefore propose a different complete case variance component test. We also derive an all case variance component test for extreme sampling designs and compare $y$-extreme and $z$-extreme sampling.

\subsection{The collapsing test}
Let $\mathbf{w}$ be a $\nu$-vector of chosen weights for the $\nu$ genetic variants in the region of interest. For the situation where all $N$ individuals are genotyped, we construct a single `collapsed' genetic covariate $\bg_w$ by a weighted sum over all $\nu$ variants $\bg_w = \G \mathbf{w}$, where $\G$ is the $N \times \nu$ genetic covariate matrix as defined in model \eqref{eq:lmygeneral}. We then assume a linear model 
$\mathbf{Y} = \X\bbeta + \bg_w \betag + \bm \varepsilon$,  which is simply a modified version of model \eqref{eq:lmysimple}, and test $H_0: \betag = 0$. For extreme sampling designs, where only $\G_{\mC}$ is observed, we define $\bg_{w,\mC} = \G_{\mC} \mathbf{w}$ and treat this $n$-vector as the observed genetic covariate. The tests for $H_0: \betag = 0$ from the previous section (complete case and all case) apply.

\subsection{The variance component test}
We assume that $\bbetag$ is a random vector with mean $0$ and variance $\tau^2 \Sigma$, where $\Sigma$ is a symmetric weight matrix. Then we test $H_0: \tau^2 = 0$ (which corresponds to $H_0: \bbetag = \bm 0$), using the variance component score test statistic \cite{wu2011SKAT,lee2012SKATO}
\begin{align}
Q = \mathbf{U}_{\bbetag}^{\T} \Sigma \mathbf{U}_{\bbetag},
\label{eq:varcomp}
\end{align}
where $\mathbf{U}_{\bbetag}$ is the score statistic from Definition \ref{def:scoretest}. We let $\Sigma = \W \W$ where $\W = \text{Diag}(\mathbf{w})$, corresponding to SKAT with a linear weighted kernel \cite{wu2011SKAT}. For some score statistic $\mathbf{U}_{\bbetag}$ (asymptotically multivariate normal) with mean $\bm 0$ and covariance $\Var(\mathbf{U}_{\bbetag})$, the asymptotic null distribution of $Q$ can be estimated by the distribution of $\sum_{m = 1}^{\nu} \lambda_m \chi^2_{1,m} $, where $\lambda_m$ are the eigenvalues of $\Sigma \Var(\mathbf{U}_{\bbetag})$ and $\chi^2_{1,m}$ are i.i.d $\chi^2_1$ random variables \cite{imhof1961}. In Supplementary Material Sections 4 and 5, complete and all case score statistics ($\mathbf{U}_{\mC}$, $\mathbf{U}_A$) and variances ($\Var(\mathbf{U}_{\mC})$, $\Var(\mathbf{U}_A)$) are derived for the different extreme sampling designs in the multivariate setting ($\nu > 1$ genetic variants). In Supplementary material Section 6, we show that the variance component test follows directly from these expressions.

\subsubsection{$z$-extreme sampling}
We make three observations regarding $z$-extreme sampling that follow from Observations \ref{obs:CCRS}, \ref{obs:asympt_zCC} and \ref{obs:z_extreme_all_complete} for single variant tests. Proofs are given in Supplementary Material Section 6.

\begin{observation}
\label{obs:varcompz}
Under $z$-extreme sampling, complete case analysis, the test that rejects $H_0: \tau^2 = 0$ if $Q_n > t_{\alpha}$, where $Q_n$ is the test statistic obtained by assuming that the complete case sample is a random sample and $t_{\alpha}$ is a critical value from the asymptotic null distribution of $Q_n$, is approximately equal to the test that rejects $H_0: \tau^2 = 0$ if $Q_{\mC} > t_{\alpha, \mC}$, where $Q_{\mC}$ is the complete case test statistic and $t_{\alpha, \mC}$ is a critical value from the asymptotic distribution of $Q_{\mC}$.
\end{observation}

\begin{observation}
A variance component score test for complete case analysis under $z$-extreme sampling derived directly from the residual phenotype model \eqref{eq:lmz} is valid, but conservative, for testing $H_0: \tau^2 = 0$.
\end{observation}

\begin{observation}
\label{obs:rareallcomp}
Under $z$-extreme sampling, the all case and complete case variance component score test statistics are approximately equal.
\end{observation}

Observations \ref{obs:varcompz} and \ref{obs:rareallcomp} also apply to $y$-extreme sampling for the particular case when there are no non-genetic covariates in the model ($\X = \mathbf{1}$).

\section{Simulations}
\label{section:simulations}

\subsection{Single variant tests}

We simulated two non-genetic covariates; $x_1$ from a $\text{N}(2,1)$ distribution, and $x_2$ from a $\text{Bernoulli}(0.5)$-distribution. The covariate $x_2$ may for example represent population substructure. In each of $10000$ simulations we drew a genetic covariate $\text{g}_i$ such that $\text{g}_i \sim \text{Binom}(2,p_0)$ if $x_{2,i} = 0$ and $\text{g}_i \sim \text{Binom}(2,p_1)$ if $x_{2,i} = 1$, $p_0 = 0.4$ and $p_1 = 0.1$ or $p_0 = p_1 = 0.25$. Phenotypes were independently generated from the model 
\begin{align*}
Y \sim \text{N}\left(5 + \beta_1 x_{1} + \beta_2 x_{2} + \beta_{\text{g}} \text{g}, 4^2 \right).
\end{align*}
Simulations were performed for three sample sizes $N = 2000$, $10000$ and $20000$. The parameter $\beta_{\text{g}}$ was 0 for estimates of type-1 error and otherwise chosen so that the power to detect an association in the full sample was between $80\%$ and $90\%$, $\beta_{\text{g}} = 0.45$, $ 0.21$, and $0.15$, for $N = 2000$, $10000$, and $20000$, respectively. Coefficients of non-genetic covariates were $\beta_1 = 10$ and $\beta_2 = 5$, or  $\beta_1 = 5$ and $\beta_2 = 2$, or $\beta_1 = 2$, and $\beta_2 = 1$. With these parameter choices the environmental covariates $x_{1}$ and $x_{2}$ were more important than the genetic covariate $\text{g}$ for describing the response ($R^2$ from fitting the regression model with and without the genetic covariates varied minimally). The latter choice was motivated by the assumption that environmental variables are more important for predicting a complex trait, compared to the genotype of a common genetic variant \cite{manolio2009finding}. For extreme sampling, we used estimated population quantiles $l = Y_{\left(\frac{n/2}{N}\right)}$, $u = Y_{\left(\frac{N - n/2}{N}\right)}$ or $l_z = Z_{\left(\frac{n/2}{N}\right)}$, $u_z = Z_{\left(\frac{N - n/2}{N}\right)}$ to ensure that $n$ individuals were selected for genotyping in a symmetric fashion. Ideally, $l$ and $u$ should be determined before seeing the data, but for large $N$, $l$ and $u$ are unbiased estimates of the population quantiles and estimated with low variance. Therefore, these estimates may be regarded as deterministic and we use them in place of the true population quantiles. Under $z$-extreme sampling, we use thresholds $l_z + \bx_i^{\T} \widehat{\bbeta}$ and $u_z + \bx_i^{\T} \widehat{\bbeta}$ where also $\widehat{\bbeta}$ is estimated based on the full population. 

Estimated power when $n=N/2$, $\beta_1 = 10$ and $\beta_2 = 5$ is presented in Table \ref{tab:power}. In addition to $y$-extreme and $z$-extreme sampling we estimated the power of randomly sampling $n$ individuals. In Figure 1 we illustrate the estimated power of all methods for increasing $n$ ($N = 10 000$ fixed), for $p_0 = 0.4$, $p_1 = 0.1$, and different values of $\beta_1$ and $\beta_2$. Increasing $n$ is equivalent to increasing $l$ and decreasing $u$. Figures were generated using the R package ggplot2 \cite{ggplot2}.

\paragraph{$y$-extreme sampling}
For $y$-extreme sampling, the all case test is clearly more powerful than the complete case test. For the simulation with $\beta_1 = 10$ and $\beta_2 = 5$ (Figure 1, panel A) the complete case method is in fact less powerful than the random sample. When the non-genetic covariates have a strong effect, then the extreme-$y$ individuals will be extreme due to the non-genetic covariates rather than the genetic covariate. As illustrated here, this can result in a particularly unfavorable sample with respect to detecting causal genetic variants. For lower non-genetic coefficient values, the power of $y$-extreme sampling designs improve (Figure 1, panels B and C). The extreme-$y$ individuals are then more likely to be extreme due to the genetic effect.

\paragraph{$z$-extreme sampling}
The $z$-extreme sampling design is clearly more powerful than $y$-extreme sampling and the power of this design is not impacted by the effect of the non-genetic covariates (Figure 1). Three theoretical results of Section \ref{section:EPSsingle} are illustrated by simulations; (i) complete case and all case score tests are approximately equally powerful (Observation \ref{obs:z_extreme_all_complete}), (ii) the complete case score test statistic is approximately equal to the score test statistic we obtain by assuming that the complete case sample is a random sample (Observation \ref{obs:CCRS}) and (iii) residual phenotype tests are less powerful than tests based on the original regression model when the genetic covariate is correlated with other covariates, but equal to tests based on the original regression model when the genetic covariate is not correlated with other covariates (Observation \ref{obs:asympt_zCC}). In our simulation set-up, $\text{g}$ is uncorrelated with non-genetic covariates when $p_0 = p_1 = 0.25$, and correlated with non-genetic covariates when $p_0 = 0.4$ and $p_1 = 0.1$.

Result (i) can be seen from Table \ref{tab:power}, where rows corresponding to $z$-extreme complete case and $z$-extreme all case analysis are almost identical. In Figure 1, the line corresponding to the $z$-extreme all case test is not visible as it lies directly underneath the line for the $z$-extreme complete case test.

Result (ii) is illustrated in Table \ref{tab:powerz}. By Observation \ref{obs:z_extreme_all_complete} we can either use the complete case score test, the all case score test (by result (i)), or we can use a score test for random samples (we used the function glm.scoretest in the R Statmod package \cite{statmod}), to test $H_0: \betag = 0$ under $z$-extreme sampling.

Result (iii) is illustrated in Table \ref{tab:power}. For the column $p_0 = 0.4$, $p_1 = 0.1$, the rows `$z$-extreme complete case (residual)' and `$z$-extreme all case (residual)' show lower power than the rows `$z$-extreme complete case' and `$z$-extreme all case'. For the column $p_0 = p_1 = 0.25$, all these tests are equal. The simulations in Figure 1 are performed with $p_0 > p_1$ and the methods based on the model for the residual phenotype never reach the power of the full model, even when all $N$ individuals are genotyped.

\paragraph{Binary response methods}

The simplest and most commonly approach to association testing under extreme sampling is to treat the extreme phenotypes as a binary responses \cite{van2000power}. We consider only the complete cases (Definition \ref{def:CC}), define a binary response vector based on low and high extreme phenotypes (either $y$ or $z$) and apply a logistic regression model. For the simulations in Table \ref{tab:power}, the $y$-extreme binary response model has almost no power to detect associations while the $z$-extreme binary response model is quite powerful. As the effect of non-genetic covariates decreases, the $y$-extreme binary response method improves (Figure 1). We note the behavior of the binary response methods in Figure 1. As $n$ increases the minor allele frequencies in each extreme tail become more similar, and it becomes harder to detect any differences between the two extremes. After some point, the sample size $n$ is not large enough to detect this small difference with sufficient power. At $n = N/2$ (which corresponds to Table \ref{tab:power}, $N = 10 000$), the $z$-extreme binary response model has good power for this particular simulation set-up, but power decreases as $n$ increases. 

\paragraph{Type 1 error rates}
Estimated type-1 error rates for $n = N/2$ (symmetric sampling) are presented in Table \ref{tab:t1error}. The residual phenotype tests are by design valid but conservative when $p_0 \neq p_1$. The $y$-extreme binary response test did not control the type-1 error rate when $N = 2000$ ($n = 1000$).

\subsection{Rare variant tests}
We performed a rare variant simulation study similar to Wu \textit{et al} \cite{wu2011SKAT}. For $10000$ individuals we generated a 1 Mb region of genetic information using the simulation program COSI (European population, standard out-of-Africa model) \cite{cosi}. For each of $10000$ simulations, we randomly selected $N = 5000$ individuals and a 30 kb region. Then, we randomly chose $50 \%$ or $10\%$ of variants with (observed) minor allele frequency below $3\%$ to be causal. The effect of a causal variant with minor allele frequency $m$ was set to $\betag = \pm c \cdot |\log_{10}(m)|$, otherwise $\betag = 0$. A certain percentage of the causal variants were set to have a negative effect sizes. The constant $c$ was 0.2 when $50\%$ of variants were causal and 1.2 when $10\%$ of variants were causal (chosen so that the power of the full sample was approximately $80\%$). Phenotypes were simulated from 
\begin{align*}
Y \sim \text{N}\left(5 + 1 x_{1} + 2 x_{2} + \G \bbetag, 4^2 \right),
\end{align*}
where $\G$ is a genotype matrix of all variants in the selected region. We estimated power of the collapsing method and the variance component method for increasing $n$ ($N$ fixed). As proposed by Wu \textit{et al} \cite{wu2011SKAT}, weights were $f_{\beta}(m,1,25)$ where $f_{\beta}()$ is the density of the Beta-distribution and $m$ is the minor allele frequency of the genetic variant. Power estimates are presented in Figure 2. For the collapsing methods, the function glm.scoretest (R Statmod package \cite{statmod}) was used for full sample, random sample and binary response tests, while our implementation of complete case and all case tests for single variants were used for extreme samples. For the variance component methods, the the SKAT R package \cite{SKATR} was used for full sample, random sample and binary response tests, while our implementation of complete case and all case variance component tests were used for extreme samples. We used the Davies method \cite{davies} as implemented in the R package CompQuadForm \cite{quadRpackage} to obtain $p$-values for the variance component score test. The power estimates presented in Figure 2 are quite similar to that of the single-variant simulation (Figure 1). The $z$-extreme sample is more powerful than the $y$-extreme sample. Under $z$-extreme sampling the complete case and all case tests are equally powerful, and under $y$-extreme sampling the all case test is more powerful than the complete case test. Concerning choice of method for rare variant association testing, we confirm the simulation results of Wu \textit{et al} \cite{wu2011SKAT}, namely that when many variants are causal and all effects in the same direction (Figure 2, panel A) the collapsing test is the more powerful, while when few variants are causal and effects are in opposite directions (Figure 2, panel B) the variant component test is the more powerful test. For the evaluation of type-1 error rates (significance level $2.5 \times 10^{-6}$, motivated by Lee \textit{et al} \cite{lee2012SKATO}) we performed 6.5 million simulations under the null. All methods (collapsing method, complete and all case, variance component method, complete and all case) rejected the null 8 times, which gives an estimated type-1 error rate of $1.2\times 10^{-6} ( 1.1\times 10^{-7}, 4.7\times 10^{-6})$ (Clopper-Pearson $95\%$ confidence interval).

\section{Application to data from the HUNT study}
\label{section:vo2}
We assess the performance of different designs and methods using data from the HUNT Fitness study \cite{HUNT, aspenes2011physical}. The study of genetic association with maximum oxygen uptake (VO2max) is an ongoing project that is run by co-authors Anja Bye, Einar Ryeng and Ulrik Wisløff. Here we use genome-wide observations of about $10^5$ common variants (minor allele frequency $\geq 0.05$) that are available for $N \approx 3000$ individuals. The study was designed so as not to include closely related participants. We assume that the sample is representative of the population of Nord-Trøndelag, Norway. The dataset consists of approximately 1500 men and 1500 women. There is a considerable difference in variance of VO2max between men and women, and we therefore base our analysis on a standardized VO2max variable where for each gender we subtract the mean and divide by the standard deviation. The null model is then a linear regression model with (standardized) VO2max as the response, and age and physical activity as non-genetic covariates.  

Since all $N$ individuals have been genotyped, we can compare the results of extreme sampling designs to the results of analyzing the full sample.  From full sample analysis it is known that a region on chromosome 1 is associated with VO2max. Plots of $p$-values (on $-\log10$-scale) from this region on chromosome 1 are shown in Figure 3. The full sample size is quite low and as could be expected, the smallest $p$-values do not reach genome-wide significance ($5\cdot10^{-7}$ for $10^5$ tests, family-wise error rate controlled at significance level $0.05$, using Bonferroni), yet it is assumed that the peak in $-\log10(p)$ represents a genetic association with VO2max. Here we attempt to replicate this finding using extreme phenotype sampling designs. We perform $y$-extreme and $z$-extreme sampling, $n = \frac{N}{2}$, so that approximately one quarter of individuals satisfy $y_i < l_i$ and one quarter satisfy $y_i > u_i$. The classification rules $l_i$ and $u_i$ were set using percentiles of the empirical phenotype or residual phenotype distributions. With both types of extreme sampling designs ($y$-extreme and $z$-extreme) we are to some extent able to replicate the genotype-phenotype association. The $y$-extreme sampling design has low power, both when using the all case and complete case score tests. The $z$-extreme sampling design is as expected more powerful, and the complete case and all case methods are equal (Observation \ref{obs:z_extreme_all_complete}). In fact, the $z$-extreme sampling results closely mimic the full sample results despite using only half of the available genetic information. Residual phenotype tests under $z$-extreme sampling can also be applied, and results are given in the Supplementary Material Section 7. These results are almost equal to those shown in Figure 3, i.e. from tests based on the original regression model (Observation \ref{obs:asympt_zCC}).

We summarize the genome-wide test results in QQ plots. QQ plots of the $p$-values (on $-\log10$-scale) for association tests on chromosomes 1 to 22 based on different extreme sampling methods are presented in Figure 4. In the construction of the QQ-plot we assume that the $p$-values from the score tests are exact, that the null model is true for all variants, and that all tests are independent. If so, each $p$-value is $\text{uniform}(0,1)$-distributed and the $i$th smallest $p$-value $p_{(i)}$ follows a $\text{Beta}(i,m+1-i)$-distribution where $m$ is the number of tests. Then $\E(p_{(i)}) = i/(m+i)$. The $2.5\%$ and $97.5\%$ quantiles of the distribution of the ordered $p$-values are also plotted. Deviations from the straight line can for example indicate associations with VO2max. The deviation seen from the full sample analysis is solely due to the region in chromosome 1 where the null model is rejected. As in Figure 3, we see that the $z$-extreme sampling design yields the results most similar to the full sample analysis.

Tests were performed separately for each chromosome. For the largest (chromosome 1), running times were 1.37 seconds for complete case analysis and 19.14 seconds for all case analysis. For reference, the full sample analysis (based on the function glm.scoretest in the Statmod package \cite{statmod}) took 3.34 seconds. All computations were performed using R version 3.4.0 \cite{Rsoftware} on a personal computer (MacBook Air (13”, Early 2014) with 1.7 GHz Intel Core i7-4650U and 4 MB cache).

\section{Discussion}

\paragraph{Improving power with extreme phenotype sampling}
The aim with using an extreme phenotype sampling design is to improve power to detect associations between genetic variants and the phenotype. It is assumed that extreme phenotype individuals are more informative with respect to genetic influence than the average person. Our simulations show that when other non-genetic or environmental covariates have a strong impact on the phenotype, then the classical extreme phenotype sampling design ($y$-extreme sampling) is not necessarily more powerful than a random sample. This is because the extremes are likely have a high or low phenotype value due to the non-genetic covariates, and not due to a genetic effect. To resolve this issue, we have defined an extreme sampling design where each person has an individual classification rule for being extreme or not. We use the residuals of the null model ($\mathbf{Y} = \X \bbeta + \bm \varepsilon$), where non-genetic effects have been accounted for, to define extreme phenotypes ($z$-extreme sampling). Using this design, the extreme individuals are more likely to be extreme due to a genetic impact, and this can improve power to detect phenotype-genotype associations.

\paragraph{Hypothesis testing for extreme phenotype samples}
We have focused on complete case and all case analysis of selectively genotyped samples. In the former, only the extreme individuals are analyzed. A binary response method can be applied to the complete cases \cite{van2000power} but dichotomizing a continuous response typically results in loss of power \cite{huang2007eps}. We have focused on a complete case method that directly models the distribution of the extremes \cite{huang2007eps}, extending current methods to allow for individual classification rules. With all case analysis we include information on non-genetic variables for non-extremes. It has previously been shown that when there are no non-genetic covariates in the model, complete case and all case tests under $y$-extreme sampling are equally powerful \cite{huang2007eps, derkach2015score}. When adding information on non-genetic covariates, the all case method is generally more powerful than complete case analysis for the classical $y$-extreme sampling design. For all case analysis we have restricted attention to a situation where the covariate to be tested is discrete, and potential confounders are discrete and completely observed. A more general all case model has been developed by Tao et al \cite{tao2017}, which for example can handle the issue of using principal components to control for confounding due to population structures.

For the $z$-extreme sampling design, we showed that the all case score test reduces to the complete case test. This is noteworthy because the complete case test requires far fewer model assumptions: one does not need to model the distribution of missing covariates and any potential dependencies between covariates, and we are not restricted to discrete genetic covariates and discrete confounders. This comes in addition to the fact that the $z$-extreme sampling design in itself is more powerful. In complete case analysis it is trivial to include principal components (missing for non-extremes) and gene-environment or gene-gene interaction terms. The complete case score test for $z$-extreme samples can furthermore be shown to be equal to the score test for a random sample (under the null, any individual is equally likely to be classified as extreme).

For $z$-extreme samples, we considered the possibility of using the residual phenotype as the response of in a regression model that under the alternative would only include genetic covariates (residual phenotype tests). We proved that the score tests derived for a model with independent observations (assuming test statistic $\chi^2$-distributed under the null) were also valid for the residual phenotype model.

\paragraph{Multiple variants and rare variant association tests}
In the Supplementary Material we have derived score tests for simultaneously testing $\nu$ genetic variants. A special case is the single variant tests presented in Section \ref{section:EPSsingle}. When simultaneously testing $\nu$ rare variants (Section \ref{sec:rarevar}) it is common to apply collapsing tests \cite{morgenthaler2007collapse, li2008collapse, madsen2009collapse, morris2010collapse} or variance component score tests such as the SKAT test \cite{wu2011SKAT}. The complete and all case collapsing methods considered here follow directly from single variant tests, while the complete and all case variance component tests were derived using results from the multivariate score tests. It then follows that also in the rare variant setting, complete case and all case tests are equal under $z$-extreme sampling, and that residual phenotype tests are valid, but at times conservative. Simulations showed that the $z$-extreme sampling design is, as for single variant testing, the more powerful design. The relative performance of the collapsing and variance component tests for extreme samples were as previosuly found for random sampling \cite{wu2011SKAT}.

\paragraph{Practice guideline}
The following guideline assumes that phenotype values $y_i$ and non-genetic covariates $\bx_i$ are observed for $N$ individuals, representative of the population on which to make inference, and that $n < N$ of these are to be genotyped. Step 1: Decide the null model $\mathbf{Y} = \X \bbeta + \varepsilon$ and calculate the residuals using standard statistical software. Step 2: From the empirical distribution of residuals, estimate thresholds $l_z$ and $u_z$ (for example so that approximately $n/2$ of individuals satisfy $z_i < l_z$ and $n/2$ satisfy $z_i > u_z$) and use these to classify individuals as extreme or not. Step 3: After genotyping extreme individuals, perform complete case hypothesis testing (using standard statistical software if no additional covariates (e.g. interactions effects or principal components) have been added to the model, otherwise using complete case tests as provided by us). Estimate parameters by maximizing the complete case likelihood. All extreme sample tests, as well as functions for parameter estimation, are available as an R package.

\section{Supplementary material}
In the Supplementary Material we present all details in the derivation of score tests for the general setting of simultaneously testing $\nu$ genetic variants, as well as analytical results regarding equality of tests under the $z$-extreme sampling design. The single variant tests presented in Sections 2 and 3 of this paper represent the special case when $\nu = 1$. The rare variant tests of Section 4 can also be seen as a particular implementation of these multivariate score tests. All methods are implemented in R and available as an R package at https://github.com/theabjorn/extremesampling. The vignette includes examples on how to make use of all the tests presented in this paper. 

\section*{Acknowledgments}
The Nord-Trøndelag Health Study (the HUNT study) is a collaboration between the HUNT Research Centre (Faculty of Medicine, Norwegian University of Science and Technology), the Nord-Trøndelag County Council, the Central Norway Health Authority and the Norwegian Institute of Public Health. 

We thank anonymous reviewers for valuable comments and suggestions. 

\section*{Conflict of interest}
None declared.

\bibliographystyle{wileyj}
\bibliography{epsbiblio}

\clearpage

\begin{table}[ht]
\centering

\begin{tabular}{lcc}
	\hline
  $p_0$, $p_1$ & 0.4, 0.1 & 0.25, 0.25 \\
  \hline
$N = 20 000$, $\beta_{\text{g}} = 0.15$ &  &  \\ 
\hline 
  Full sample & 0.856 (0.849,0.863) & 0.934 (0.929,0.939) \\ 
  Random sample & 0.575 (0.565,0.585) & 0.690 (0.681,0.699) \\ 
  $y$-extreme complete case & 0.462 (0.452,0.471) & 0.560 (0.550,0.569) \\ 
  $y$-extreme all case & 0.615 (0.606,0.625) & 0.732 (0.724,0.741) \\ 
  $z$-extreme complete case & 0.830 (0.823,0.838) & 0.914 (0.909,0.920) \\ 
  $z$-extreme complete case (residual) & 0.766 (0.757,0.774) & 0.915 (0.909,0.920) \\ 
  $z$-extreme all case & 0.830 (0.823,0.838) & 0.914 (0.909,0.920) \\ 
  $z$-extreme all case (residual) & 0.766 (0.758,0.774) & 0.914 (0.909,0.920) \\ 
  $y$-extreme binary & 0.118 (0.111,0.124) & 0.131 (0.125,0.138) \\ 
  $z$-extreme binary & 0.695 (0.686,0.704) & 0.873 (0.866,0.879) \\ 
   \hline
$N = 10 000$, $\beta_{\text{g}} = 0.21$ &  &  \\ 
\hline 
  Full sample & 0.856 (0.849,0.863) & 0.894 (0.888,0.900) \\ 
  Random sample & 0.567 (0.557,0.576) & 0.626 (0.617,0.636) \\ 
  $y$-extreme complete case & 0.451 (0.441,0.461) & 0.500 (0.491,0.510) \\ 
  $y$-extreme all case & 0.610 (0.601,0.620) & 0.662 (0.653,0.672) \\ 
  $z$-extreme complete case & 0.833 (0.826,0.840) & 0.868 (0.861,0.874) \\ 
  $z$-extreme complete case (residual) & 0.760 (0.752,0.769) & 0.868 (0.861,0.874) \\ 
  $z$-extreme all case & 0.833 (0.826,0.841) & 0.868 (0.861,0.875) \\ 
  $z$-extreme all case (residual) & 0.760 (0.751,0.768) & 0.868 (0.861,0.874) \\ 
  $y$-extreme binary & 0.111 (0.105,0.117) & 0.114 (0.108,0.120) \\ 
  $z$-extreme binary & 0.690 (0.680,0.699) & 0.821 (0.814,0.829) \\ 
   \hline
$N = 2 000$, $\beta_{\text{g}} = 0.45$ &  &  \\ 
\hline   
  Full sample & 0.827 (0.819,0.834) & 0.869 (0.863,0.876) \\ 
  Random sample & 0.527 (0.517,0.537) & 0.586 (0.577,0.596) \\ 
  $y$-extreme complete case & 0.416 (0.407,0.426) & 0.475 (0.465,0.485) \\ 
  $y$-extreme all case & 0.576 (0.567,0.586) & 0.640 (0.631,0.650) \\ 
  $z$-extreme complete case & 0.799 (0.791,0.807) & 0.850 (0.843,0.857) \\ 
  $z$-extreme complete case (residual) & 0.722 (0.713,0.731) & 0.850 (0.843,0.857) \\ 
  $z$-extreme all case & 0.797 (0.789,0.805) & 0.850 (0.843,0.857) \\ 
  $z$-extreme all case (residual) & 0.721 (0.712,0.730) & 0.850 (0.843,0.857) \\ 
  $y$-extreme binary & 0.113 (0.107,0.120) & 0.119 (0.113,0.125) \\ 
  $z$-extreme binary & 0.658 (0.649,0.667) & 0.792 (0.784,0.800) \\ 
   \hline
\end{tabular}
\caption{Estimated power (with Clopper-Pearson $95\%$ confidence intervals) for single variant tests at significance level $\alpha = 0.05$ and with sample size $n = N/2$; random sampling, $y$-extreme sampling and $z$-extreme sampling. Residual phenotype tests under the $z$-extreme sampling design are denoted by (residual). Coefficients $\beta_0 = 5$, $\beta_{1} = 10$, $\beta_2 = 5$, $\betag = 0$.}
\label{tab:power}
\end{table}

\begin{table}[ht]
\centering

\begin{tabular}{lcc}
	\hline
  $p_0$, $p_1$ & 0.4, 0.1 & 0.25, 0.25 \\
  \hline
$N = 20 000$, $\beta_{\text{g}} = 0.15$ &  &  \\ 
\hline 
$z$-extreme complete case (random) & 0.836 (0.828,0.843) & 0.881 (0.874,0.887) \\ 
  $z$-extreme complete case & 0.836 (0.829,0.843) & 0.881 (0.874,0.887) \\ 
  $z$-extreme all case & 0.836 (0.829,0.843) & 0.880 (0.874,0.887) \\
   \hline
   $N = 20 000$, $\beta_{\text{g}} = 0$  &  &  \\ 
\hline 
$z$-extreme complete case (random) & 0.049 (0.045,0.054) & 0.051 (0.047,0.056) \\ 
  $z$-extreme complete case & 0.049 (0.045,0.054) & 0.051 (0.047,0.056) \\ 
  $z$-extreme all case & 0.049 (0.045,0.054) & 0.051 (0.047,0.056) \\ 
  \hline
$N = 10 000$, $\beta_{\text{g}} = 0.21$ &  &  \\ 
\hline 
$z$-extreme complete case (random) & 0.825 (0.818,0.833) & 0.871 (0.865,0.878) \\ 
  $z$-extreme complete case & 0.825 (0.818,0.833) & 0.872 (0.865,0.878) \\ 
  $z$-extreme all case & 0.825 (0.818,0.833) & 0.870 (0.864,0.877) \\ 
   \hline
      $N = 10 000$, $\beta_{\text{g}} = 0$ &  &  \\ 
\hline 
$z$-extreme complete case (random) & 0.052 (0.048,0.056) & 0.051 (0.047,0.056) \\ 
  $z$-extreme complete case & 0.052 (0.048,0.057) & 0.051 (0.047,0.056) \\ 
  $z$-extreme all case & 0.052 (0.048,0.057) & 0.051 (0.047,0.056) \\
  \hline
$N = 2 000$, $\beta_{\text{g}} = 0.45$ &  &  \\ 
\hline   
$z$-extreme complete case (random) & 0.796 (0.788,0.804) & 0.842 (0.835,0.849) \\ 
  $z$-extreme complete case & 0.797 (0.789,0.805) & 0.843 (0.835,0.850) \\ 
  $z$-extreme all case & 0.796 (0.788,0.804) & 0.842 (0.834,0.849) \\ 
   \hline
   $N = 2 000$, $\beta_{\text{g}} = 0$ &  &  \\ 
\hline 
$z$-extreme complete case (random) & 0.048 (0.044,0.052) & 0.047 (0.043,0.051) \\ 
  $z$-extreme complete case & 0.048 (0.044,0.053) & 0.047 (0.043,0.052) \\ 
  $z$-extreme all case & 0.048 (0.044,0.053) & 0.047 (0.043,0.051) \\ 
  \hline
\end{tabular}
\caption{Estimated power ($\betag > 0$) and type-1 error rates ($\betag = 0$) (with Clopper-Pearson $95\%$ confidence intervals) for $z$-extreme sampling and single variant tests at significance level $\alpha = 0.05$, applying a score test for random samples, a complete case test with $l_i$ and $u_i$ as defined for $z$-extreme sampling, and applying the all case score test. Sample size $n = N/2$. Coefficients $\beta_0 = 5$, $\beta_{1} = 10$, $\beta_2 = 5$.}
\label{tab:powerz}
\end{table}

\begin{table}[ht]
\centering

\begin{tabular}{lcc}
	\hline
  $p_0$, $p_1$ & 0.4, 0.1 & 0.25, 0.25 \\
  \hline
$N = 20 000$, $\beta_{\text{g}} = 0$ &  &  \\ 
\hline 
Full sample & 0.048 (0.044,0.052) & 0.052 (0.048,0.057) \\ 
  Random sample & 0.046 (0.042,0.051) & 0.052 (0.047,0.056) \\ 
  $y$-extreme complete case & 0.048 (0.044,0.053) & 0.050 (0.046,0.054) \\ 
  $y$-extreme all case & 0.048 (0.044,0.052) & 0.051 (0.047,0.055) \\ 
  $z$-extreme complete case & 0.048 (0.044,0.053) & 0.052 (0.048,0.056) \\ 
  $z$-extreme complete case (residual) & 0.028 (0.025,0.031) & 0.052 (0.048,0.056) \\ 
  $z$-extreme all case & 0.048 (0.044,0.053) & 0.052 (0.048,0.056) \\ 
  $z$-extreme all case (residual) & 0.028 (0.025,0.031) & 0.052 (0.048,0.056) \\ 
  $y$-extreme binary & 0.050 (0.046,0.055) & 0.052 (0.048,0.057) \\ 
  $z$-extreme binary & 0.030 (0.027,0.033) & 0.051 (0.047,0.056) \\ 
   \hline
$N = 10 000$, $\beta_{\text{g}} = 0$ &  &  \\ 
\hline 
Full sample & 0.049 (0.045,0.053) & 0.052 (0.047,0.056) \\ 
  Random sample & 0.051 (0.047,0.056) & 0.051 (0.047,0.056) \\ 
  $y$-extreme complete case & 0.051 (0.046,0.055) & 0.052 (0.048,0.056) \\ 
  $y$-extreme all case & 0.052 (0.048,0.056) & 0.051 (0.046,0.055) \\ 
  $z$-extreme complete case & 0.049 (0.045,0.053) & 0.051 (0.047,0.055) \\ 
  $z$-extreme complete case (residual) & 0.027 (0.024,0.031) & 0.051 (0.047,0.056) \\ 
  $z$-extreme all case & 0.049 (0.045,0.053) & 0.051 (0.047,0.056) \\ 
  $z$-extreme all case (residual) & 0.027 (0.024,0.031) & 0.051 (0.047,0.056) \\ 
  $y$-extreme binary & 0.054 (0.049,0.059) & 0.050 (0.046,0.055) \\ 
  $z$-extreme binary & 0.031 (0.028,0.034) & 0.050 (0.046,0.054) \\ 
   \hline
$N = 2 000$, $\beta_{\text{g}} = 0$ &  &  \\ 
\hline   
Full sample & 0.050 (0.045,0.054) & 0.051 (0.046,0.055) \\ 
  Random sample & 0.050 (0.046,0.055) & 0.050 (0.046,0.055) \\ 
  $y$-extreme complete case & 0.053 (0.049,0.058) & 0.049 (0.045,0.054) \\ 
  $y$-extreme all case & 0.052 (0.048,0.057) & 0.048 (0.044,0.052) \\ 
  $z$-extreme complete case & 0.049 (0.045,0.054) & 0.051 (0.046,0.055) \\ 
  $z$-extreme complete case (residual) & 0.027 (0.024,0.030) & 0.050 (0.046,0.055) \\ 
  $z$-extreme all case & 0.049 (0.045,0.053) & 0.050 (0.046,0.054) \\ 
  $z$-extreme all case (residual) & 0.027 (0.024,0.030) & 0.050 (0.046,0.055) \\ 
  $y$-extreme binary & 0.063 (0.058,0.068) & 0.056 (0.051,0.061) \\ 
  $z$-extreme binary & 0.030 (0.026,0.033) & 0.050 (0.046,0.054) \\ 
   \hline
\end{tabular}
\caption{Estimated type-1 error rates (with Clopper-Pearson $95\%$ confidence intervals) for single variant tests at significance level $\alpha = 0.05$ and with sample size $n = N/2$; random sampling, $y$-extreme sampling and $z$-extreme sampling. Residual phenotype tests under the $z$-extreme sampling design are denoted by (residual). Coefficients $\beta_0 = 5$, $\beta_{1} = 10$, $\beta_2 = 5$, $\betag = 0$.}
\label{tab:t1error}
\end{table}

\clearpage

\begin{figure}[ht]
    \begin{center}
       \includegraphics[width=5in]{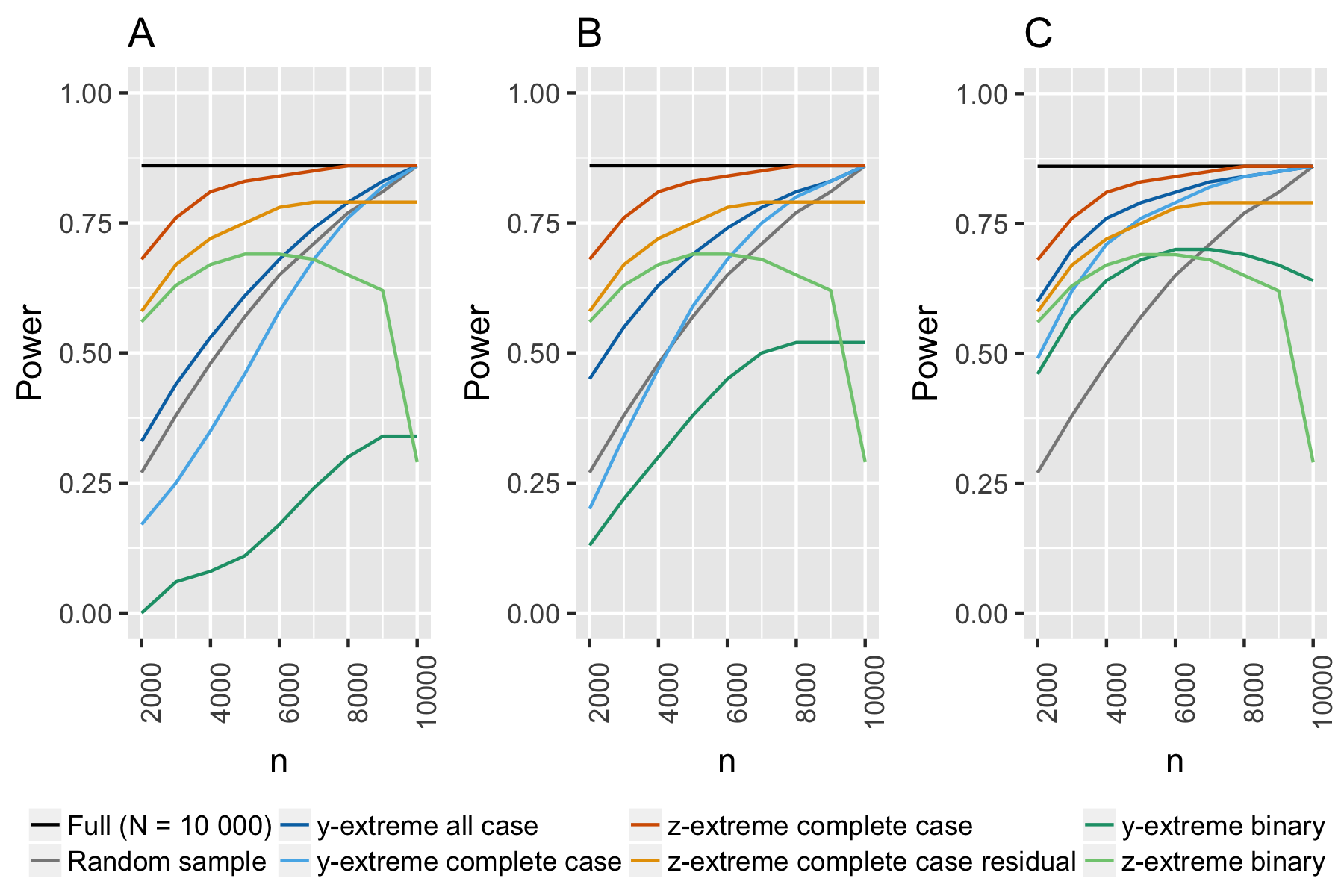}
    \end{center}
\end{figure}

\paragraph{Figure 1:} 
Single variant tests. Estimated power (significance level $\alpha = 0.05$) of different designs (random, $y$-extreme and $z$-extreme sampling) and methods (complete case, all case and binary response) as the number of genotyped individuals ($n$) approaches the full sample size ($N = 10000$). For any $n$, the $n/2$ individuals with the lowest values of $y$ (or $z$) and the $n/2$ individuals with the highest values of $y$ (or $z$) were classified as extreme and therefore had observed genotypes. A: $\beta_{1} = 10$, $\beta_{2} = 5$, B:  $\beta_{1} = 5$, $\beta_{2} = 2$,  C:  $\beta_{1} = 2$, $\beta_{2} = 1$. All simulations: $p_0 = 0.4, p_1 = 0.1$, $\betag = 0.21$.

\clearpage

\begin{figure}[ht]
    \begin{center}
       \includegraphics[width=5in]{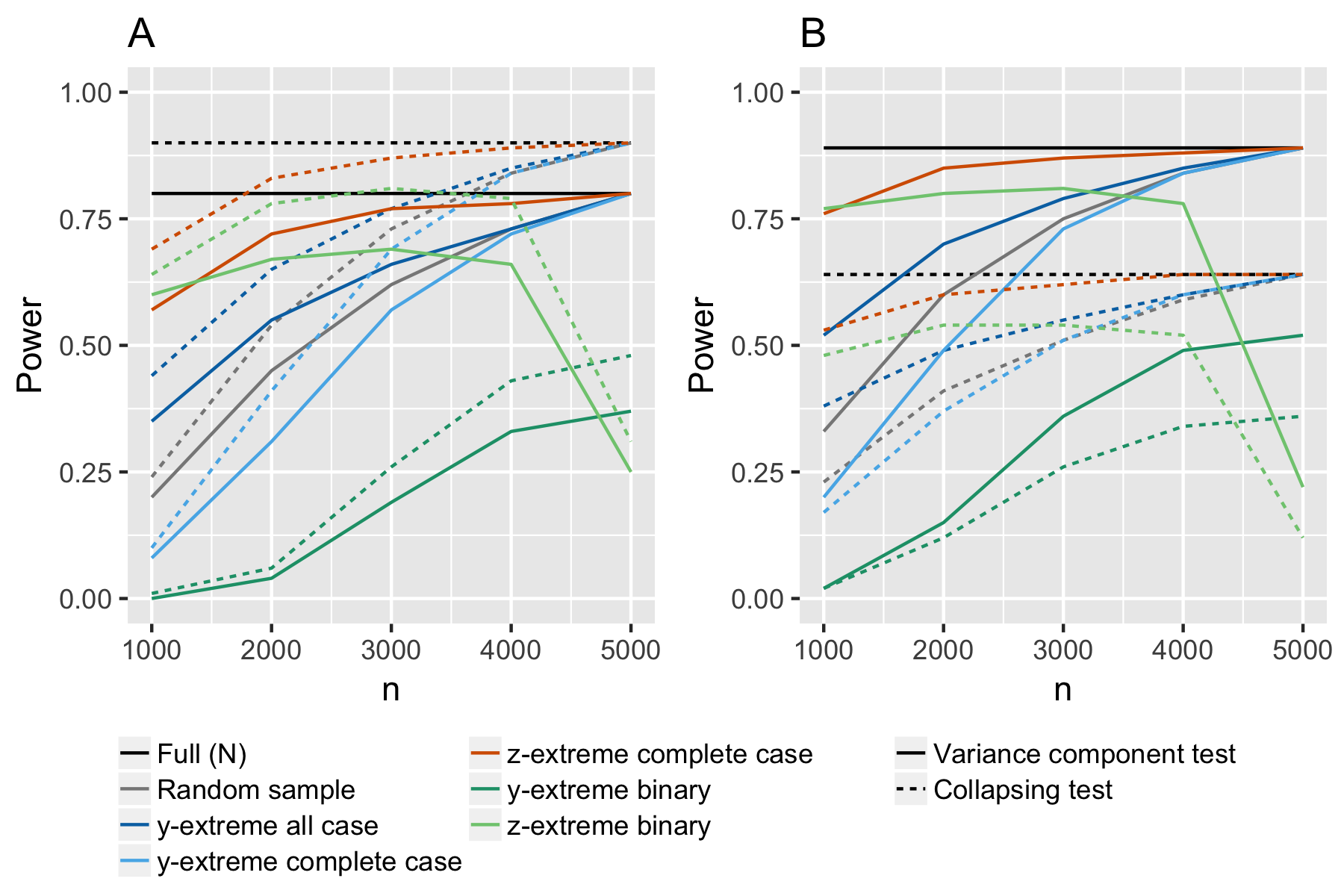}
    \end{center}
\end{figure}

\paragraph{Figure 2:} 
Rare variant tests. Estimated power (significance level $\alpha = 2.5\times 10^{-6}$) of different designs (random, $y$-extreme and $z$-extreme sampling) and methods (complete case, all case and binary response) as the number of genotyped individuals ($n$) approaches the full sample size ($N$). For any $n$, the $n/2$ individuals with the lowest values of $y$ (or $z$) and the $n/2$ individuals with the highest values of $y$ (or $z$) were classified as extreme, and had observed genotypes. A: $N=5000$ and $50\%$ of variants with minor allele frequency below $3\%$ in a 30kb region were causal, all had effect in the positive direction, $|c| = 0.2$. B: $N=5000$ and $10\%$ of variants with minor allele frequency below $3\%$ in a 30kb region were causal, $80\%$ of these had effect in the positive direction and $20\%$ in the negative direction, $|c| = 1.2$. For the $z$-extreme sampling design the complete case residual phenotype test, all case test and all case residual phenotype test are equal to the complete case test and therefore not shown.

\clearpage

\begin{figure}[ht]
    \begin{center}
       \includegraphics[width=5in]{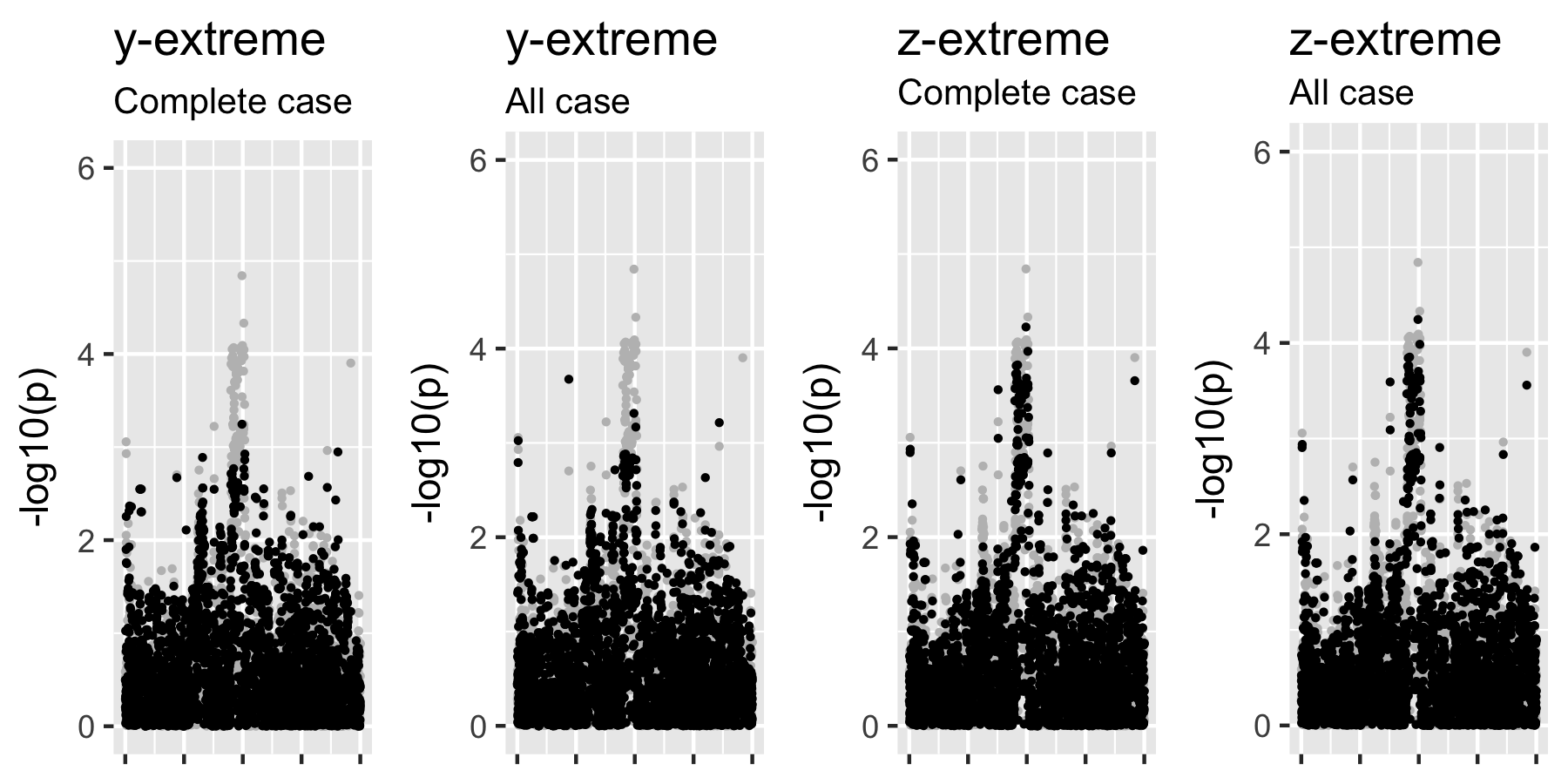}
    \end{center}
\end{figure}

\paragraph{Figure 3:} 
Association tests for a selected region on chromosome 1. The extreme samples are based on $n=N/2$ extreme individuals (symmetric sampling). The results of the full sample is shown as shaded points.

\begin{figure}[ht]
    \begin{center}
       \includegraphics[width=5in]{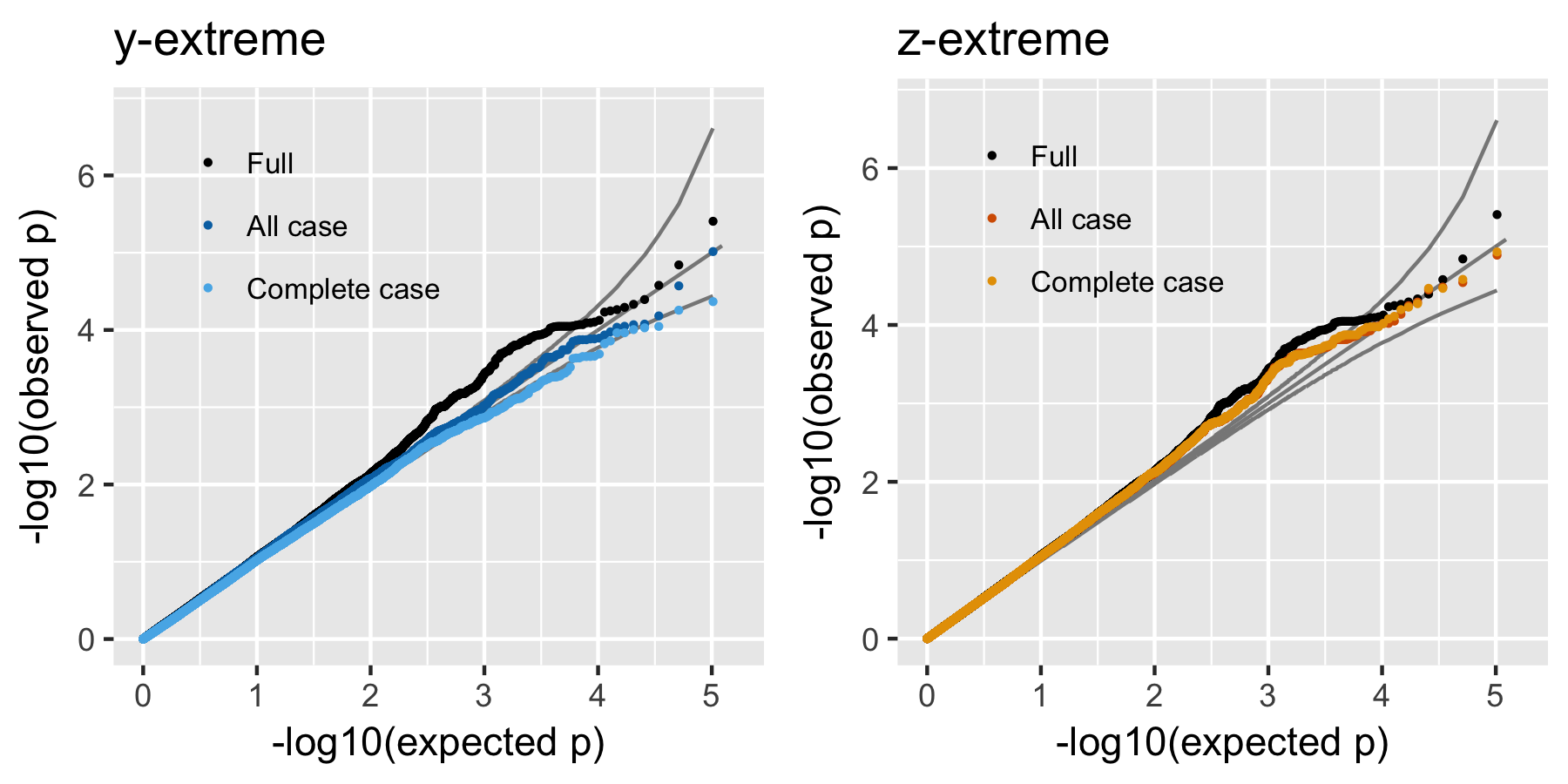}
    \end{center}
\end{figure}

\paragraph{Figure 4:} 
Quantile-quantile plots of $p$-values for genome-wide association testing in the VO2max study. The extreme samples are based on $n=N/2$ extreme individuals (symmetric sampling).

\end{document}